 \documentclass[12pt]{article}
 \usepackage[cp1251]{inputenc}
 \usepackage[english, russian]{babel}
 \usepackage{amsmath, latexsym, amssymb, bm, array, graphics,
 amsfonts}
 \makeatletter

\newcommand{\si}{\mathop{\rm sign\; \mu}}
 \makeatother
 
\usepackage[dvips]{graphicx}

\begin{document}
\thispagestyle{empty}
\large
\renewcommand{\abstractname}{Abstract}
\renewcommand{\refname}{\begin{center}
 REFERENCES\end{center}}
\makeatother

\begin{center}
\bf  Boundary problems for the one-dimensional kinetic equation with
the collisional frequence proportional to the module velocity of
molecules
\end{center} \medskip

\begin{center}
  \bf

  A. V. Latyshev\footnote{$avlatyshev@mail.ru$}
  A. A. Yushkanov\footnote{$yushkanov@inbox.ru$}
\end{center}\medskip

\begin{center}
{\it Faculty of Physics and Mathematics,\\ Moscow State Regional
University, 105005,\\ Moscow, Radio str., 10A}
\end{center}\medskip

\begin{abstract}
For the one-dimensional linear kinetic
equations with  collisional frequency of the molecules,
proportional to the module
velocity of molecules, analytical solutions
of problems about
temperature jump and weak evaporation (con\-den\-sation) in
rarefied gas are received .
Quantities of  temperature and concentration
jumps are found. Distributions of concentration, mass velocity and
temperature are constructed. Necessary numerical calculations
and graphic researches are done.

{\bf Key words:} kinetic equation, collisional frequency,
boundary problems, analytical solution, distribution of
macroparameters.
\medskip

PACS numbers:  05.60.-k   Transport processes,
51.10.+y   Kinetic and transport theory of gases,

\end{abstract}

\begin{center}
\bf   Introduction
\end{center}
In work \cite{1} the linear one-dimensional kinetic  equation
with collisional integral BGK (Bhatnagar, Gross and
Krook) and  collisional frequency, affine depending on the module
of velocity of molecules has been entered.
Preservation laws numerical density (concentration) of molecules,
mo\-men\-tum and energy of molecules thus have been  used.

In \cite{1} the theorem about structure of common solution
of the entered equation has been proved.

In work \cite{2}, being by continuation \cite {1}, are received
exact solutions of a problem on temperature jump and weak evaporation
(condensation) in the rarefied gas. These two problems following
\cite{3} we will name the generalized problem of Smolukhovsky.

In work \cite{44} the generalized problem of Smolukhovsky has been solved
for special case of affine dependence of collisional frequency
on the module of velocity of molecules.
Namely, the case, when collisional fre\-qu\-ency is constant has been
considered.

In the present work other limiting case affine dependence when
frequ\-en\-cy of collisions is proportional to the module of velocity
of molecules is considered.
On the basis of the analytical solution of Smolukhovsky' boundary
problem the function of distribution of gas molecules is
const\-ruc\-ted.
Numerical values of  jump of temperature and
weak evaporation (condensation) coefficients  are found.
Distributions of concentration, mass velocity and temperature in
"half-space"\, are constructed.

Let us stop on history of exclusively analytical solutions
of the gene\-ra\-li\-zed Smolukhovsky' problem.

For the simple (one-atomic) rarefied gas with constant
frequency of collisions of molecules the analytical
solution of the generalized
Smolu\-khov\-sky' problem was received in \cite{4}.

In \cite{5} the generalized Smolukhovsky'  problem was analytically
solved for the simple rarefied gas with collisional frequency
of molecules, linearly depending on the module
velocity of molecules.

In \cite{6} the problem about strong evaporation (condensation)
with constant frequency of collisions has been analytically solved.

Let us notice, that for the first time the problem about
temperature jump with collisional frequency of molecules,
linearly depending on the module of molecular velocity, was
analytically solved by Cassell and Williams in work \cite{7} in
1972.

Then in works \cite{8,9,10} the  Smolukhovsky' problem has been
genera\-lized on case of multiatomic (molecular) gases and also
the analytical solution was received.

In works \cite{11,12,13} the problem close to the problem
about tempera\-ture jump for electrons, about behaviour of the quantum
Bose-gas at low temperatures  is considered. It has been thus used
the kinetic equation with fonons excitation agrees N.N. Bogolyubov.

In works \cite{14,15} the problem about temperature jump
for electrons of degenerate plasmas in metal has been solved.

In work \cite{16} the analytical solution of Smolukhovsky' problem
and for quantum gases  has been received.

In work of Cercignani and Frezzotti \cite{17} the Smolukhovsky'
problem was considered with use of the one-dimensional kinetic
equations. The complete analytical solution of the  Smolukhovsky'
problem with  use of Cercignani---Frezzotti equation
has been received in work \cite{18}.

In the present work the analytical solution of
the generalized Smoluk\-hov\-sky'   problem is considered.
The case of collisional frequency propor\-tional to the module of
molecular velocity, in model of one-dimensional gas is considered.
Model of one-dimensional gas
gave the good consent with the results devoted to the three-dimensional
gas \cite{18}.

\begin{center}
  {\bf 1. Statement of the  problem and basic equations}
\end{center}

Let us start  with statement of the  Smolukhovsky' problem for
the one-dimensional kinetic equation with frequency of collisions,
affinne  de\-pen\-ding on the module  of molecular velocity.

Let gas occupies half-space $x>0$.
The surface temperature $T_s$ and
concentration of sated steam of a surface $n_s$ are set.
Far from the surface at $x=0$ gas moves with some velocity $u$,
being velocity of evaporation (or con\-den\-sation),
also has the temperature gradient
$$
g_T=\Big(\dfrac{d\ln T}{dx}\Big)_{x=+\infty}.
$$

It is necessary to define
temperature and concentration jumps depen\-ding on velocity $u$ and
temperature gradient $g_T$.

In the problem about weak evaporation (condensation)
is required to define temperature and concentration jumps
depending on velocity, including a temperature gradient
equal to zero, and velocity of evapora\-tion (condensation) is enough small.
The last means, that
$$
u \ll v_T.
$$

Here $v_T$ is the themal velocity of molecules, having an order
of a sound velocity,
$$
v_T=\dfrac{1}{\sqrt{\beta_s}}, \qquad \beta_s=\dfrac{m}{2k_BT_s},
$$
$m$ is the mass of molecule, $k_B$ is the Boltzmann constant.

In the problem about temperature jump is required to define
tempera\-ture and concentration jumps
depending on a temperature gradient, thus velocity
evaporation (condensation)
is considered equal to zero, and the temperature gradient is
considered small.
It means, that
$$
lg_T\ll 1, \qquad l=\tau v_T,\qquad \tau=\dfrac{1}{\nu_0},
$$
where $l$ is the mean free path of gas molecules,
$\tau$ is the time of relaxation, i.e. time between two
consecutive collisions of molecules.

Let us unite both problems (about weak evaporation (condensation) and
temperature jump) in one. We will assume a smallness of gradient
temperature (i.e. a smallness of relative difference
temperature on length of free puth) and a smallness
velocity of gas in comparison with velocity of a sound. In this
case the problem supposes linearization and function of distribution
it is possible to search for  in the form
$$
f(x,v)=f_0(v)(1+h(x,v)),
$$
where
$$
f_0(v)=n_s\Big(\dfrac{m}{2\pi k_BT_s}\Big)^{1/2}
\exp \Big[-\dfrac{mv^2}{2k_BT_s}\Big]
$$
is the absolute Maxwellian.

Let us pass to dimensionless velocity
$$
\mu=\sqrt{\beta}v=\dfrac{v}{v_T}
$$
and  dimensionless coordinate
$$
x'=\nu_0 \sqrt{\dfrac{m}{2k_BT_s}}x=\dfrac{x}{l}
$$

Variable $x'$ we will designate again through $x$.

We take the kinetic equation \cite{1}
$$
\mu\dfrac{\partial h}{\partial x}+(1+\sqrt{\pi}a|\mu|)h(x,\mu)=
$$
$$
=(1+\sqrt{\pi}a|\mu|)\dfrac{1}{\sqrt{\pi}}\int\limits_{-\infty}^{\infty}
e^{-\mu'^2}(1+\sqrt{\pi}a|\mu'|)q(\mu,\mu',a)h(x,\mu')d\mu'.
\eqno{(1.1)}
$$

Here $q(\mu,\mu',a)$ is the kernel of equation, \medskip
$$
q(\mu,\mu',a)=r_0(a)+r_1(a)\mu\mu'+r_2(a)(\mu^2-\beta(a))(\mu'^2-\beta(a)),
$$ \medskip
$$
r_0(a)=\dfrac{1}{a+1},\qquad r_1(a)=\dfrac{2}{2a+1},\qquad
r_2(a)=\dfrac{4(a+1)}{4a^2+7a+2},
$$ \medskip
$$
\beta=\beta(a)=\dfrac{2a+1}{2(a+1)}.
$$

At $a\to 0$ from the equation (1.1) we receive the following
kinetic equation with constant frequency of collisions
$$
\mu\dfrac{\partial h}{\partial x}+h(x,\mu)=$$$$=\dfrac{1}{\sqrt{\pi}}
\int\limits_{-\infty}^{\infty}e^{\mu'^2}
\Big[1+2\mu\mu'+2\Big(\mu^2-\dfrac{1}{2}\Big)
\Big(\mu'^2-\dfrac{1}{2}\Big)\Big]h(x,\mu')d\mu'.
$$

Let us deduce the kinetic equation in a second limiting case,
when collisional frequency is proportional to the module
of the molecular velocity.

Let us consider the second limiting case of the equation (1.1).
We will return to expression of frequency of collisions also
we will copy it in the form
$$
\nu(\mu)=\nu_0(1+\sqrt{\pi}a|\mu|)=\nu_0+\nu_1|\mu|,
$$
where
$$
\nu_1=\sqrt{\pi}\nu_0 a.
$$
Let us $\nu_0$ tends to zero. In this limit the quantity $a $ tends to
$ + \infty $, for
$$
a =\dfrac {\nu_1} {\sqrt {\pi} \nu_0}.
$$
It is easy to see, that in this limit
$$
\lim\limits_{a\to+\infty}(1+\sqrt{\pi}a|\mu'|)q(\mu,\mu',a)=
\sqrt{\pi}|\mu'|q_1(\mu,\mu'),
$$
where
$$
q_1(\mu,\mu')=1+\mu\mu'+(\mu^2-1)(\mu'^2-1).
$$

Thus the equation (1.1) will  be  transformed in the form
$$
\dfrac{\mu}{|\mu|}\dfrac{\partial h}{\partial x_1}+h(x_1,\mu)=
\int\limits_{-\infty}^{\infty}e^{-\mu'^2}|\mu'|q_1(\mu,\mu')h(x_1,\mu)d\mu'.
\eqno{(1.2)}
$$

In this equation
$$
x_1=\nu_1\sqrt{\beta_s}x=\dfrac{x}{l_1},\qquad l_1=v_T\tau_1,\qquad
\tau_1=\dfrac{1}{\nu_1}.
$$

This equation is the one-dimensional kinetic equation with
colli\-sio\-nal frequency proportional to the module of the molecular
velocity.

The equation (1.2) it is possible to present in the form
$$
\si\dfrac{\partial h}{\partial x_1}+h(x_1,\mu)=
\int\limits_{-\infty}^{\infty}e^{-\mu'^2}|\mu'|q_1(\mu,\mu')h(x_1,\mu)d\mu'.
\eqno{(1.2')}
$$

The equation $ (1.2')$ contains two equations. One of these
equations
$$
+\dfrac{\partial h}{\partial x_1}+h(x_1,\mu)=
\int\limits_{-\infty}^{\infty}e^{-\mu'^2}|\mu'|q_1(\mu,\mu')h(x_1,\mu)d\mu'
\eqno{(1.2'')}
$$
is defined in the phase quarter-planes $\mathbb{R}^+ = \{(x, \mu):
x> 0, \mu> 0 \} $, and another equation
$$
-\dfrac{\partial h}{\partial x_1}+h(x_1,\mu)=
\int\limits_{-\infty}^{\infty}e^{-\mu'^2}|\mu'|q_1(\mu,\mu')h(x_1,\mu)d\mu'
\eqno{(1.2''')}
$$
is defined in other phase quarter-planes $\mathbb{R}^-=\{(x,\mu):
x>0, \mu<0\}$.

Let us solve further the generalized  Smolukhovsky' problem for the equation
(1.2). At first it is required to formulate correctly the generalized
Smolukhovsky' problem  as a boundary problem of mathema\-ti\-cal physics.

\begin{center}
  \bf 2. The kinetic equation with collisional frequency
proportional to the module of molecular velocity,
and the problem statement
\end{center}

Rectilinear substitution it is possible to check up, that the kinetic
equation (1.2) has following four partial solutions
$$
h_0(x,\mu)=1,
$$
$$
h_1(x,\mu)=\mu,
$$
$$
h_2(x,\mu)=\mu^2,
$$
$$
h_3(x,\mu)=\Big(\mu^2-\dfrac{3}{2}\Big)(x-{\rm sign\;}\mu).
$$

Let us consider, that molecules are reflected from a wall purely
dif\-fu\-sively, i.e. are reflected from a wall with Maxwell distribution on
velocity, i.e.
$$
f(x,v)=f_0(v),\qquad v_x>0.
$$
From here for function $h (x, \mu) $ we receive
$$
h(0,\mu)=0, \qquad \mu>0.
\eqno{(2.1)}
$$

Condition (2.1) is the first boundary condition to the equation (1.2).

For asymptotics  Chapmen---Enskog distribution we will search in
the form of the linear combination of its partial solutions
with unknown coef\-fi\-ci\-ents
$$
h_{as}(x,\mu)=A_0+A_1\mu+A_2\Big(\mu^2-\dfrac{1}{2}\Big)+$$$$+
A_3\Big[\Big(\mu^2-\dfrac{3}{2}\Big)(x-\si)-\dfrac{1}{\sqrt{\pi}}\mu\Big].
\eqno{(2.2)}
$$

Let us notice, that in (2.2) velocity mode is orthogonal to
thermal mode, i.e.
$$
\int\limits_{-\infty}^{\infty}e^{-\mu^2}\mu
\Big[\Big(\mu^2-\dfrac{3}{2}\Big)(x-\si)-
\dfrac{1}{\sqrt{\pi}}\mu\Big]d\mu=0.
$$
Besides,  constant mode is orthogonal to  temperature mode
$$
\int\limits_{-\infty}^{\infty}e^{-\mu^2}\Big(\mu^2-\dfrac{1}{2}\Big)
d\mu=0.
$$

For definition of four constants $A_0, A_1, A_2, A_3$
let us take advantage of definitions of macroparameters of gas:
concentration, mass velocity, temperature and jumps of temperature and
concentration (numerical density).

Let us consider distribution of numerical density
$$
n(x)=\int\limits_{-\infty}^{\infty}f(x,v)dv=
\int\limits_{-\infty}^{\infty}f_0(v)(1+h(x,v))dv=
n_0+\delta n(x).
$$

Here
$$
n_0=\int\limits_{-\infty}^{\infty}f_0(v)dv,\qquad
\delta n(x)=\int\limits_{-\infty}^{\infty}f_0(v)h(x,v)dv.
$$
From here we find that
$$
\dfrac{\delta n(x)}{n_0}=\dfrac{1}{\sqrt{\pi}}
\int\limits_{-\infty}^{\infty}e^{-\mu^2}h(x,\mu)d\mu.
$$

We denote
$$
n_e=n_0\dfrac{1}{\sqrt{\pi}}
\int\limits_{-\infty}^{\infty}e^{-\mu^2}(1+h_{as}(x=0,\mu))d\mu.
$$

From here we receive that
$$
\varepsilon_n\equiv \dfrac{n_e-n_0}{n_0}=\dfrac{1}{\sqrt{\pi}}
\int\limits_{-\infty}^{\infty}e^{-\mu^2}h_{as}(x=0,\mu)d\mu.
\eqno{(2.3)}
$$

The quantity $\varepsilon_n$ is the required quantity of concentration
jump.

Substituting (2.2) in (2.3), we find that
$$
\varepsilon_n=A_0.
\eqno{(2.4)}
$$

From definition of dimensional velocity of gas
$$
u(x)=\dfrac{1}{n(x)}\int\limits_{-\infty}^{\infty}f(x,v)vdv
$$
we receive that in linear approximation dimensional mass velocity
equals
$$
U(x)=\dfrac{1}{\sqrt{\pi}}\int\limits_{-\infty}^{\infty}
e^{-\mu^2}h(x,\mu)\mu d\mu.
$$

Setting "far from a wall" \, velocity of evaporation (condensation),
let us write
$$
U=\dfrac{1}{\sqrt{\pi}}\int\limits_{-\infty}^{\infty}
e^{-\mu^2}h_{as}(x,\mu)\mu d\mu.
\eqno{(2.5)}
$$

Substituting in (2.5) distribution (2.2), we receive, that
$$
U=\dfrac{\sqrt{\pi}}{2}A_1.
\eqno{(2.6)}
$$

We consider the temperature distribution
$$
T(x)=\dfrac{2}{kn(x)}\int\limits_{-\infty}^{\infty}\dfrac{m}{2}
(v-u_0(x))^2f(x,v)dv.
$$

From here we find that
$$
\dfrac{\delta T(x)}{T_0}=-\dfrac{\delta n(x)}{n_0}+\dfrac{2}{\sqrt{\pi}}
\int\limits_{-\infty}^{\infty}e^{-\mu^2}h(x,\mu)\mu^2d\mu=
$$
$$
=\dfrac{2}{\sqrt{\pi}}
\int\limits_{-\infty}^{\infty}e^{-\mu^2}h(x,\mu)(\mu^2-\dfrac{1}{2})d\mu.
$$

Now from here follows, that at $x\to + \infty $ asymptotic
distribution is equal
$$
\dfrac{\delta T_{as}(x)}{T_0}=\dfrac{2}{\sqrt{\pi}}
\int\limits_{-\infty}^{\infty}e^{-\mu^2}h_{as}(x,\mu)
(\mu^2-\dfrac{1}{2})d\mu.
\eqno{(2.7)}
$$

Definition of a gradient of temperature far from a wall means, that
distribution of temperature looks like
$$
T(x)=T_e+\Big(\dfrac{dT}{dx}\Big)_{x=+\infty}\cdot x=T_e+G_Tx,
$$
where
$$
G_T=\Big(\dfrac{dT}{dx}\Big)_{+\infty}.
$$

This distribution we will present in the form
$$
T(x)=T_s\Big(\dfrac{T_e}{T_s}+g_Tx\Big)=T_s\Big(1+
\dfrac{T_e-T_s}{T_s}+g_Tx\Big), \quad x\to +\infty,
$$
where
$$
g_T=\Big(\dfrac{d\ln T}{dx}\Big)_{x=+\infty},
$$
or
$$
T(x)=T_s(1+\varepsilon_T+g_Tx),\qquad x\to +\infty,
\eqno{(2.8)}
$$
where
$$
\varepsilon_T=\dfrac{T_e-T_s}{T_s}
$$
is the required quantity of temperature jump.

From expression (2.8) it is visible, that relative temperature change
far from a wall is described by linear function
$$
\dfrac{\delta T_{as}(x)}{T_s}=\dfrac{T(x)-T_s}{T_s}=
\varepsilon_T+g_Tx,\quad x\to+\infty.
\eqno{(2.9)}
$$

Substituting (2.2) in (2.7), we receive, that
$$
\dfrac{\delta T_{as}(x)}{T_s}=A_2+A_3x.
\eqno{(2.10)}
$$

Comparing (2.9) and (2.10), we find
$$
A_2=\varepsilon_T, \qquad A_3=g_T.
$$

So, asymptotic distribution function of Chapmen---Ensfog
is const\-ruc\-ted
$$
h_{as}(x,\mu)=\varepsilon_n+
(2U-g_T)\dfrac{\mu}{\sqrt{\pi}}+$$$$+
\varepsilon_T\Big(\mu^2-\dfrac{1}{2}\Big)+g_T
\Big(\mu^2-\dfrac{3}{2}\Big)(x-\si).
\eqno{(2.11)}
$$

Now we will formulate the second boundary condition to the equation (1.2)
$$
h(x,\mu)=h_{as}(x,\mu)+o(1), \qquad x\to +\infty.
\eqno{(2.12)}
$$

Now we will formulate the basic boundary problem, named "gene\-ra\-lized
Smolukhovsky' problem". This problem consists in  finding of the such
solution of the kinetic equation (1.2), which satisfies
to boundary conditions (2.1) and (2.12), and in (2.12)
asymptotic function of Chap\-men---Enskog distribution $h_{as}(x, \mu) $
is defined by equality (2.11).

\begin{center}
  \bf 3. The general solution of the one-dimensional kinetic equation
\end{center}

Let  us notice, that the continuous spectrum of the characteristic equation,
answering to the initial equation (1.2), represents the empty
set (see \cite{2} and \cite{44}). This fact speaks that
the equation (1.2) does not contain convection
derivative.

Therefore for the solution of the initial equation (1.2)
we will search in the form of a polynom on  velocity variable.
We search for the solution in the form
its linear combination of invariants collisions with unknown
coefficients depending from "spatial" \,  variable
$$
h(x,\mu)=a_0(x)+a_1(x)\mu+a_2(x)(\mu^2-1)+\hspace{3cm}
$$
$$
\hspace{4cm}+\si [b_0(x)+b_1(x)\mu+b_2(x)(\mu^2-1)].
\eqno{(3.1)}
$$

Distribution  function (3.1) contains two distribution functions.
One function
$$
h^+(x,\mu)=a_0(x)+a_1(x)\mu+a_2(x)(\mu^2-1)+
$$
$$
+b_0(x)+b_1(x)\mu+b_2(x)(\mu^2-1), \quad \mu>0.
\eqno{(3.1')}
$$
describes the molecules flying to the wall.

The second function
$$
h^-(x,\mu)=a_0(x)+a_1(x)\mu+a_2(x)(\mu^2-1)-$$$$-
[b_0(x)+b_1(x)\mu+b_2(x)(\mu^2-1)], \quad \mu<0.
\eqno{(3.1'')}
$$
describes the molecules reflected from a wall.

The left part of the equation (1.2) is equal to the sum of expressions
$$
\si \dfrac{\partial h^{\pm}}{\partial x}=\Bigg\{
\begin{array}{c}
  a_0'+b_0'+(a_1'+b_1')\mu+(a_2'+b_2')(\mu^2-1), \quad \mu>0, \\
  -(a_0'-b_0')- (a_1'-b_1')\mu-(a_2'-b_2')(\mu^2-1), \quad \mu<0,
\end{array}
$$
and
$$
h^{\pm}(x,\mu)=(a_0\pm b_0)+(a_1\pm b_1)\mu+(a_2\pm b_2)(\mu^2-1),\quad
\pm \mu>0.
$$

Thus, at $ \mu> 0$ the left part of the equation (1.2) for
the molecules reflected from a wall is equal
$$
\dfrac{\partial h^+}{\partial x}+h^+(x,\mu)=[a_0(x)+b_0(x)+a_0'(x)+
b_0'(x)]+$$$$+\mu [a_1(x)+b_1(x)+a_1'(x)+b_1'(x)]+$$$$+
(\mu^2-1)[a_2(x)+b_2(x)+a_2'(x)+b_2'(x)].
$$

At $ \mu <0$ the right part of the equation (1.2) for flying to a wall
molecules is equal
$$
-\dfrac{\partial h^-}{\partial x}+h^-(x,\mu)=[a_0(x)-b_0(x)-a_0'(x)+
b_0'(x)]+$$$$+\mu [a_1(x)-b_1(x)-a_1'(x)+b_1'(x)]+$$$$+
(\mu^2-1)[a_2(x)-b_2(x)-a_2'(x)+b_2'(x)].
$$

The right part of the equation (1.2) is equal
$$
R[h(x,\mu)]=\int\limits_{0}^{\infty}e^{-\mu'^2}\mu'
q(\mu,\mu')h^+(x,\mu')d\mu'+
$$
$$
+\int\limits_{0}^{\infty}e^{-\mu'^2}\mu' q(\mu,-\mu')h^-(x,-\mu')d\mu'.
$$

Let us substitute in this right part distribution function
$$
h^{+}(x,\mu)=(a_0+b_0)+(a_1+ b_1)\mu+(a_2+b_2)(\mu^2-1)
$$
and
$$
h^{-}(x,-\mu)=(a_0-b_0)-(a_1-b_1)\mu+(a_2-b_2)(\mu^2-1).
$$

We receive, that the right part is equal
$$
R[h]=\dfrac{1}{2}[a_0(x)+b_0(x)]+\dfrac{\sqrt{\pi}}{4}[a_1(x)+
b_1(x)]+
$$
$$
+\mu\left\{\dfrac{\sqrt{\pi}}{4}[a_0(x)+b_0(x)]+\dfrac{1}{2}[a_1(x)+
b_1(x)]+\dfrac{\sqrt{\pi}}{8}[a_2(x)+b_2(x)]\right\}+
$$
$$
+(\mu^2-1)\left\{\dfrac{\sqrt{\pi}}{8}[a_1(x)+b_1(x)]+
\dfrac{1}{2}[a_2(x)+b_2(x)]\right\}+
$$
$$
+\dfrac{1}{2}[a_0(x)-b_0(x)]-\dfrac{\sqrt{\pi}}{4}[a_1(x)-
b_1(x)]+
$$
$$
-\mu\left\{\dfrac{\sqrt{\pi}}{4}[a_0(x)-b_0(x)]-\dfrac{1}{2}[a_1(x)-
b_1(x)]+\dfrac{\sqrt{\pi}}{8}[a_2(x)-b_2(x)]\right\}+
$$
$$
+(\mu^2-1)\left\{-\dfrac{\sqrt{\pi}}{8}[a_1(x)-b_1(x)]+
\dfrac{1}{2}[a_2(x)-b_2(x)]\right\}.
$$

Let us simplify the previous expression
$$
R[h]=a_0(x)+\dfrac{\sqrt{\pi}}{2}b_1(x)+\mu
\Big[\dfrac{\sqrt{\pi}}{2}b_0(x)+a_1(x)+
\dfrac{\sqrt{\pi}}{4}b_2(x)\Big]+$$$$+(\mu^2-1)
\Big[\dfrac{\sqrt{\pi}}{4}b_1(x)+a_2(x)\Big].
$$

Let us equate the left and right parts of the equation (1.2).
We will receive system, consisting of six equations
$$
a_0'+b_0'+b_0=\dfrac{\sqrt{\pi}}{4}b_1,
$$
$$
a_0'-b_0'+b_0=-\dfrac{\sqrt{\pi}}{4}b_1,
$$
$$
a_1'+b_1'+b_1=\dfrac{\sqrt{\pi}}{2}b_0+\dfrac{\sqrt{\pi}}{4}b_2,
$$
$$
-a_1'+b_1'-b_1=\dfrac{\sqrt{\pi}}{2}b_0+\dfrac{\sqrt{\pi}}{4}b_2,
$$
$$
a_2'+b_2'+b_2=\dfrac{\sqrt{\pi}}{4}b_1,
$$
$$
-a_2'+b_2'-b_2=\dfrac{\sqrt{\pi}}{4}b_1.
$$

Adding the first equation with the second, the third with the
fourth, the fifth with the sixth, and then subtracting,
we will simplify this system
$$
a_0'(x)+b_0(x)=0,
\eqno{(3.2)}
$$
$$
b_0'(x)=\dfrac{\sqrt{\pi}}{2}b_1(x),
\eqno{(3.3)}
$$
$$
b_1'(x)=\dfrac{\sqrt{\pi}}{2}b_0(x)+\dfrac{\sqrt{\pi}}{4}b_2(x),
\eqno{(3.4)}
$$
$$
a_1'(x)+b_1(x)=0,
\eqno{(3.5)}
$$
$$
b_2'(x)=\dfrac{\sqrt{\pi}}{4}b_1(x),
\eqno{(3.6)}
$$
$$
a_2'(x)+b_2(x)=0.
\eqno{(3.7)}
$$

We differentiate the equation (3.4) and we will take advantage
of the equations (3.3) and (3.6). We receive the equation
$$
b_1''(x)=\dfrac{5\pi}{16}b_1(x),
$$
whence we find
$$
b_1(x)=B_1e^{-\gamma_0x},
\eqno{(3.8)}
$$
where $B_1$ is the arbitrary constant, and
$$
\gamma_0=\dfrac{\sqrt{5\pi}}{4}\approx 0.9908.
$$

From the equations (3.6) and (3.3) by means of (3.8) we receive
$$
b_2(x)=-\dfrac{2}{\sqrt{5}}B_1e^{-\gamma_0x}+B_2,
\eqno{(3.9)}
$$
$$
b_0(x)=-\dfrac{2}{\sqrt{5}}B_1e^{-\gamma_0x}+B_0,
\eqno{(3.10)}
$$
where $B_0$ and $B_2$ are arbitrary constants.

From the equation (3.5) by means of (3.8) it is found
$$
a_1(x)=-\dfrac{1}{\gamma_0}B_1e^{-\gamma_0x}+A_1,
\eqno{(3.11)}
$$
where $A_1$ is the arbitrary constant.

From the equation (3.2) by means of (3.11) it is found
$$
a_0(x)=-\dfrac{8}{5\sqrt{\pi}}B_1e^{-\gamma_0x}-B_0x+A_0,
\eqno{(3.12)}
$$
where $A_0$ is the arbitrary constant.

At last, from the equation (3.7) by means of (3.9) it is found
$$
a_2(x)=-\dfrac{8}{5\sqrt{\pi}}B_1e^{-\gamma_0x}-B_2x+A_2,
\eqno{(3.13)}
$$
where $A_2$ is the arbitrary constant.

Let us write out on the basis of equalities (3.8) -- (3.13)
general solution of equation (1.2) in the explicit form
$$
h(x,\mu)=-\dfrac{2}{\sqrt{5}}B_1e^{-\gamma_0x}-B_0x+A_0+\mu
\Big[-\dfrac{1}{\gamma_0}B_1e^{-\gamma_0x}+A_1\Big]+
$$
$$
+(\mu^2-1)\Big[-\dfrac{8}{5\sqrt{\pi}}B_1e^{-\gamma_0x}-B_2x+A_2\Big]+
\si \Big\{-\dfrac{2}{\sqrt{5}}B_1e^{-\gamma_0x}+B_0+
$$
$$
+\mu B_1e^{-\gamma_0x}+(\mu^2-1)
\Big[-\dfrac{2}{\sqrt{5}}B_1e^{-\gamma_0x}+B_2\Big]\Big\}.
\eqno{(3.14)}
$$

Let us allocate in this decision (3.14) exponential decreasing and
polynomial solutions
$$
h(x,\mu)=B_1e^{-\gamma_0x}
\Big(\mu-\dfrac{\sqrt{\pi}}{2\gamma_0}\mu^2\Big)
\Big(\dfrac{1}{\gamma_0}+\si\Big)+
$$
$$
+A_0+A_1\mu+A_2(\mu^2-1)+(\si -x)[B_0+B_2(\mu^2-1)].
\eqno{(3.15)}
$$

\begin{center}
  \bf 4. The solution of the generalized Smolukhovsky' problem
\end{center}

In this item we will prove the theorem about the analytical
solution of the basic boundary problem
(1.2), (2.1) and (2.11).

{\sc Theorem.} \;{\it The boundary problem (1.2), (2.1) and (2.11) has
the unique solution, representable in the form of the sum
exponential decreasing and polynomial solutions
$$
h(x,\mu)=-(2U-g_T)\dfrac{e^{-\gamma_0x}}{\sqrt{\pi}}
\dfrac{1+\gamma_0\si}{1+\gamma_0}
\Big(\mu-\dfrac{\sqrt{\pi}}{2\gamma_0}\mu^2\Big)+
$$
$$
+\varepsilon_n+\varepsilon_T+
(2U-g_T)\dfrac{\mu}{\sqrt{\pi}}+\Big(\mu^2-\dfrac{3}{2}\Big)
[\varepsilon_T+g_T(x-\si)],
\eqno{(4.1)}
$$
and quantities of temperature jump $ \varepsilon_T $ and
concentration jump $ \varepsilon_n $ are given by equalities}
$$
\varepsilon_T=\Big(1+\dfrac{1}{2\gamma_0}\Big)g_T-
\dfrac{1}{2\gamma_0}(2U),
\eqno{(4.2)}
$$
and
$$
\varepsilon_n=-\Big(1-\dfrac{1}{4\gamma_0}\Big)g_T-\dfrac{1}{4\gamma_0}
(2U).
\eqno{(4.3)}
$$

The solution (4.1) contains solutions of two problems: problem about
temperature jump (when $U=0$) (see fig. 1)
$$
\dfrac{h^T(x,\mu)}{g_T}=\dfrac{e^{-\gamma_0x}}{\sqrt{\pi}}
\dfrac{1+\gamma_0\si}{1+\gamma_0}\Big(\mu-\dfrac{2\mu^2}{\sqrt{5}}\Big)-
$$
$$
-\Big(1-\dfrac{1}{4\gamma_0}\Big)-\dfrac{\mu}{\sqrt{\pi}}+
\Big(1+\dfrac{1}{\gamma_0}\Big)\Big(\mu^2-\dfrac{1}{2}\Big)+
(x-\si)\Big(\mu^2-\dfrac{3}{2}\Big),
$$
and problem about weak evaporation (whenа $g_T=0$) (see fig. 2)
$$
\dfrac{h^U(x,\mu)}{2U}=-\dfrac{e^{-\gamma_0x}}{\sqrt{\pi}}
\dfrac{1+\gamma_0\si}{1+\gamma_0}\Big(\mu-\dfrac{2\mu^2}{\sqrt{5}}\Big)+
\dfrac{1}{4\gamma_0}+\dfrac{\mu}{\sqrt{\pi}}-
\dfrac{\mu^2}{2\gamma_0}.
$$

{\sc Proof.}
Let us take advantage of boundary condition "far from a wall" \,
(2.12). We will substitute in  condition (2.12) the decomposition
(3.15). We receive following equation
$$
A_0+A_1\mu+A_2(\mu^2-1)-(x-\si) [B_0+B_2(\mu^2-1)]=
$$
$$
=\varepsilon_n+(2U-g_T)\dfrac{\mu}{\sqrt{\pi}}+
\varepsilon_T\Big(\mu^2-\dfrac{1}{2}\Big)+
g_T\Big(\mu^2-\dfrac{3}{2}\Big)(x-\si).
$$

From here at once we  find
$$
A_1=\dfrac{2U}{\sqrt{\pi}}-\dfrac{g_T}{\sqrt{\pi}},
$$
$$
B_2=-g_T,
$$
$$
B_0=\dfrac{g_T}{2},
$$
$$
A_2=\varepsilon_T,
$$
$$
A_0=\varepsilon_n+\dfrac{\varepsilon_T}{2}.
$$

Let us take advantage of the boundary condition reflexion
of molecules from a wall.
Let us substitute decomposition (3.15) in the boundary condition
(2.1). We receive the algebraic equation
$$
B_1\Big(\mu-\dfrac{2}{\sqrt{5}}\mu^2\Big)\Big(\dfrac{1}{\gamma_0}+1\Big)
+\varepsilon_n+$$$$+\dfrac{2U-g_T}{\sqrt{\pi}}\mu+\varepsilon_T
\Big(\mu^2-\dfrac{1}{2}\Big)-g_T\Big(\mu^2-\dfrac{3}{2}\Big)=0.
$$

From here we receive system from three equations
$$
\varepsilon_n-\dfrac{\varepsilon_T}{2}+\dfrac{3}{2}g_T=0,
$$
$$
B_1\Big(1+\dfrac{1}{\gamma_0}\Big)+(2U-g_T)\dfrac{1}{\sqrt{\pi}}=0,
$$
$$
-B_1\dfrac{2}{\sqrt{5}}\Big(1+\dfrac{1}{\gamma_0}\Big)+
\varepsilon_T-g_T=0.
$$

From these equations we find the constant $B_1$
$$
B_1=-\dfrac{2U-g_T}{\sqrt{\pi}(1+1/\gamma_0)},
$$
and also quantities of temperature and concentration jumps:
$$
\varepsilon_T=\Big(1+\dfrac{1}{2\gamma_0}\Big)g_T-
\dfrac{1}{2\gamma_0}(2U),
\eqno{(4.4)}
$$\medskip
$$
\varepsilon_n=-\Big(1-\dfrac{1}{4\gamma_0}\Big)g_T-
\dfrac{1}{4\gamma_0}(2U).
\eqno{(4.5)}
$$\medskip

Formulas (4.4) and (4.5) in accuracy coincide with formulas of
tem\-pe\-ratures and concentration  jumps (4.2) and (4.3).

Thus, the solution of the boundary problem is constructed and has
the following form
$$
h(x,\mu)=-\dfrac{(2U-g_T)e^{-\gamma_0x}}
{\sqrt{\pi}\Big(1+\dfrac{1}{\gamma_0}\Big)}
(\mu-\dfrac{2\mu^2}{\sqrt{5}})\Big(\si
+\dfrac{1}{\gamma_0}\Big)+
$$
$$
+\varepsilon_n+(2U-g_T)\dfrac{\mu}{\sqrt{\pi}}+
\varepsilon_T\Big(\mu^2-\dfrac{1}{2}\Big)+
(x-\si)\Big(\mu^2-\dfrac{3}{2}\Big)g_T.
\eqno{(4.6)}
$$

Expansion (4.6) in accuracy coincides with expansion (4.1),
if to consider expressions (4.2) and (4.3) for quantities of
temperatures and concentration jumps.
The theorem is proved. \medskip

Let us notice, that polynomial "tail" \, of solution (4.1) is
asymptotic Chapmen---Enskog expansion. It means,
that solution (4.1) it is possible to present in the form
$$
h(x,\mu)=-\dfrac{(2U-g_T)e^{-\gamma_0x}}
{\sqrt{\pi}\Big(1+\dfrac{1}{\gamma_0}\Big)}h^*(\mu)+h_{as}(x,\mu),
$$
where
$$
h^*(\mu)=(\mu-\dfrac{2\mu^2}{\sqrt{5}})\Big(\si
+\dfrac{1}{\gamma_0}\Big).
$$

Let us transform Chapmen---Enskog decomposition
$$
h_{as}(x,\mu)=$$$$=\varepsilon_n+
(2U-g_T)\dfrac{\mu}{\sqrt{\pi}}+
\varepsilon_T\Big(\mu^2-\dfrac{1}{2}\Big)+g_T
\Big(\mu^2-\dfrac{3}{2}\Big)(x-\si)
$$
by means of equalities for temperature and concentration jumps.
As result we receive, that
$$
h_{as}(x,\mu)=$$$$=g_T\Big[-\Big(1-\dfrac{1}{4\gamma_0}\Big)-
\dfrac{\mu}{\sqrt{\pi}}+\Big(1+\dfrac{1}{2\gamma_0}\Big)\Big(\mu^2
-\dfrac{1}{2}\Big)+(x-\si)\Big(\mu^2-\dfrac{3}{2}\Big)\Big]+
$$
$$
+(2U)\Big[\dfrac{1}{4\gamma_0}+\dfrac{\mu}{\sqrt{\pi}}-
\dfrac{\mu^2}{2\gamma_0}\Big].
\eqno{(4.7)}
$$\medskip

Thus, definitively distribution function in the generalized
Smolukhov\-sky' problem is equal
$$
h(x,\mu)=-\dfrac{e^{-\gamma_0x}}{\sqrt{\pi}}(2U-g_T)\Big(\mu-
\dfrac{2\mu^2}{\sqrt{5}}\Big)\dfrac{1+\gamma_0\si}{1+\gamma_0}
+h_{as}(x,\mu),
\eqno{(4.8)}
$$
and Chapmen---Enskog decomposition $h_{as}(x, \mu) $ is defined
by equality (4.7).

Expansion (4.8) in accuracy coincides with expansion (4.1).

{\sc Remark 4.1.}
Expression (4.8) contains distribution function of  ref\-lec\-ted
molecules from wall:
$$
h^+(x,\mu)=-\dfrac{e^{-\gamma_0x}}{\sqrt{\pi}}(2U-g_T)
\Big(\mu-\dfrac{2\mu^2}{\sqrt{5}}\Big)+
$$
$$
+\varepsilon_n+\varepsilon_T+\dfrac{2U-g_T}{\sqrt{\pi}}\mu+
\Big(\mu^2-\dfrac{3}{2}\Big)[\varepsilon_T+g_T(x-1)].
$$
and also distribution function of molecules flying to the wall, which
it is expressed by Chapmen---Enskog distribution
$$
h^-(x,\mu)=-\dfrac{e^{-\gamma_0x}}{\sqrt{\pi}}(2U-g_T)
\dfrac{1-\gamma_0}{1+\gamma_0}\Big(\mu-\dfrac{2\mu^2}{\sqrt{5}}
\Big)+
$$
$$
+\varepsilon_n+\varepsilon_T+\dfrac{2U-g_T}{\sqrt{\pi}}\mu+
\Big(\mu^2-\dfrac{3}{2}\Big)[\varepsilon_T+g_T(x+1)].
$$
\medskip

\begin{figure}[h]
\begin{center}
\includegraphics[width=14cm,height=14cm]{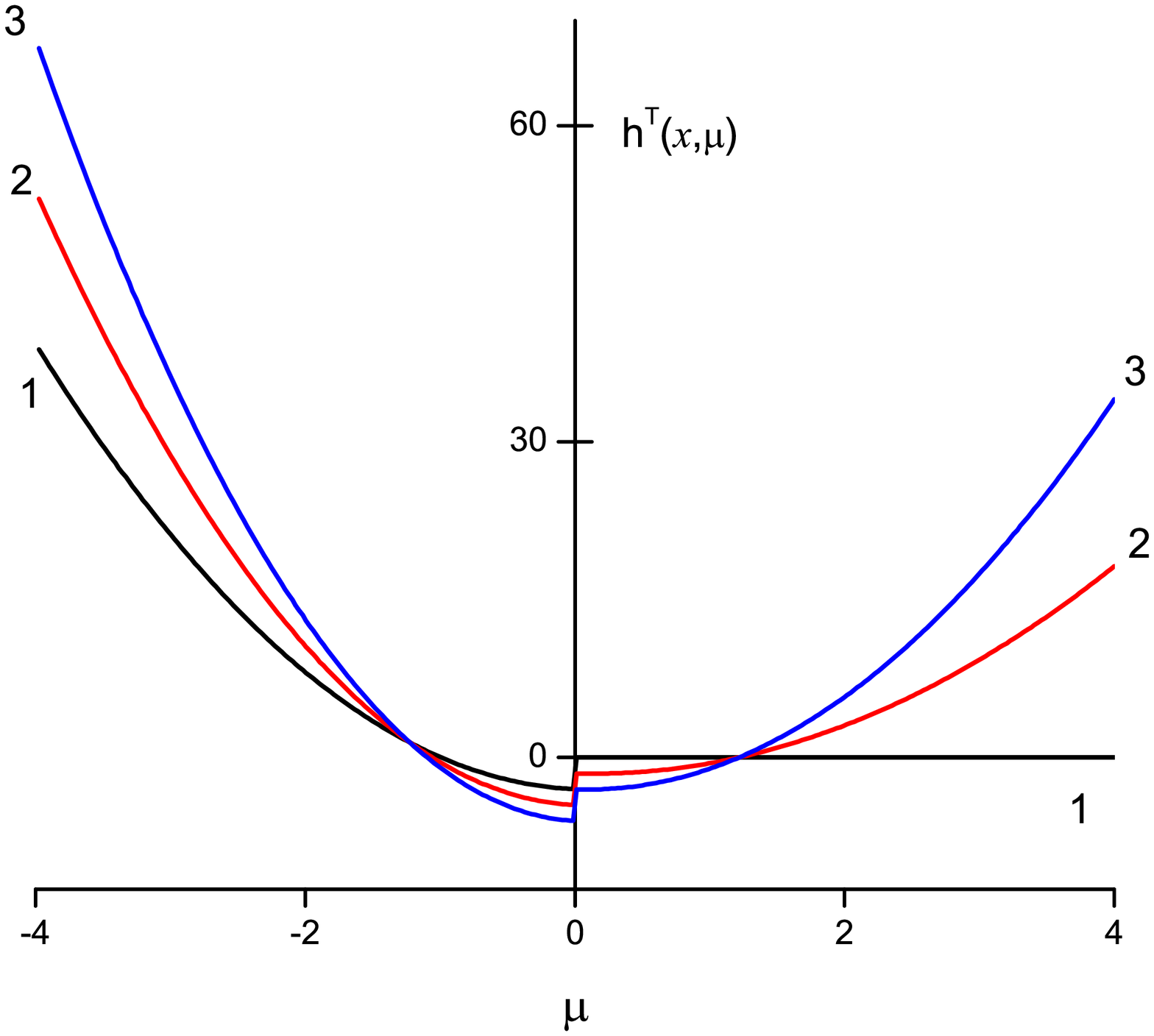}
\end{center}
\begin{center}
{Fig. 1. Distribution function in problem about temperature jump.
Curves $1,2,3$ correspond to values $x=0,1,2$.}
\end{center}
\end{figure}

\begin{figure}[t]
\begin{center}
\includegraphics[width=14cm,height=14cm]{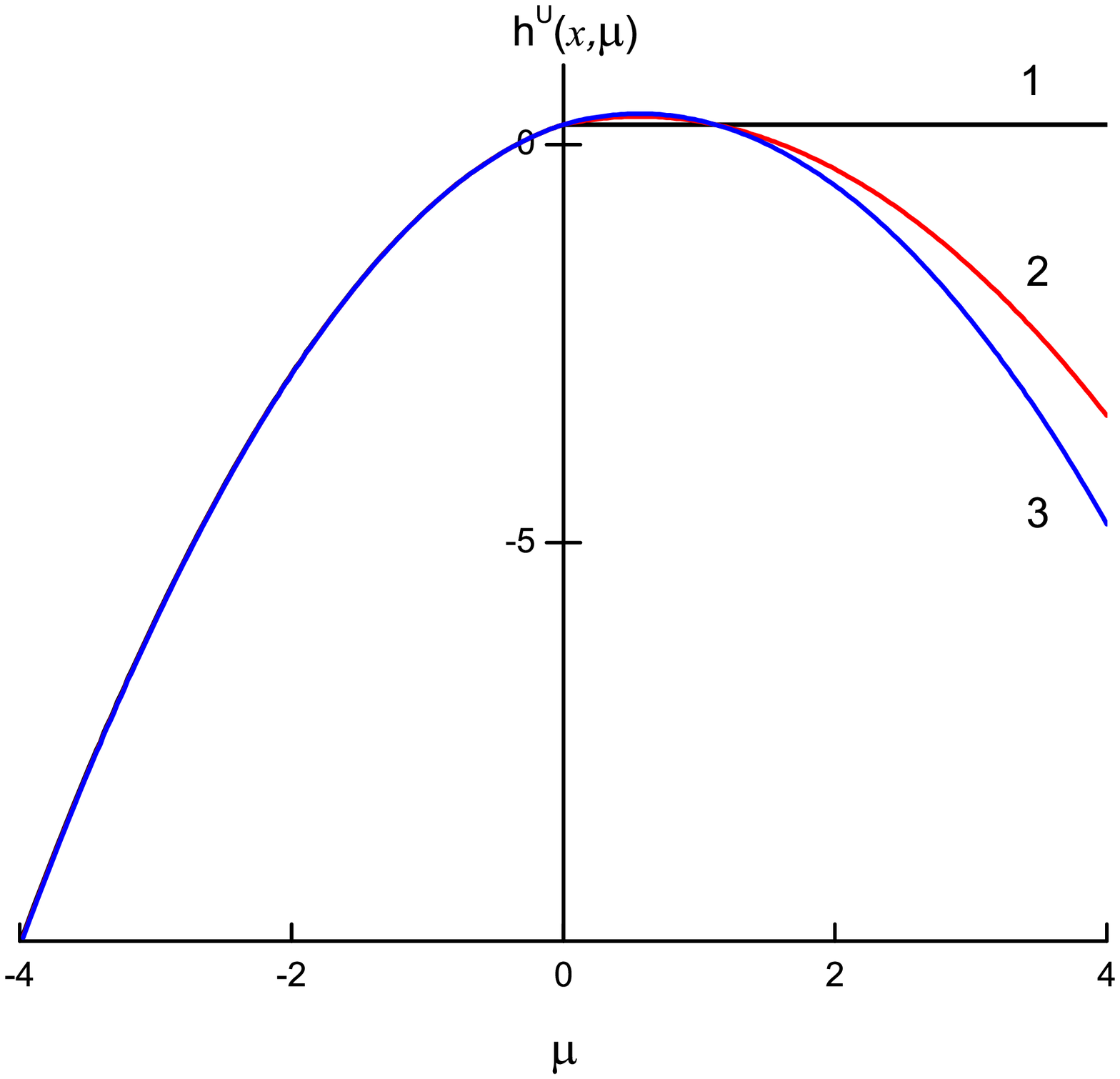}
\end{center}
\begin{center}
{Fig. 2. Distribution function in problem about weak evaporation.
Curves $1,2,3$ correspond to values $x=0,1,2$.}
\end{center}
\end{figure}

\begin{figure}[t]
\begin{center}
\includegraphics[width=16cm,height=14cm]{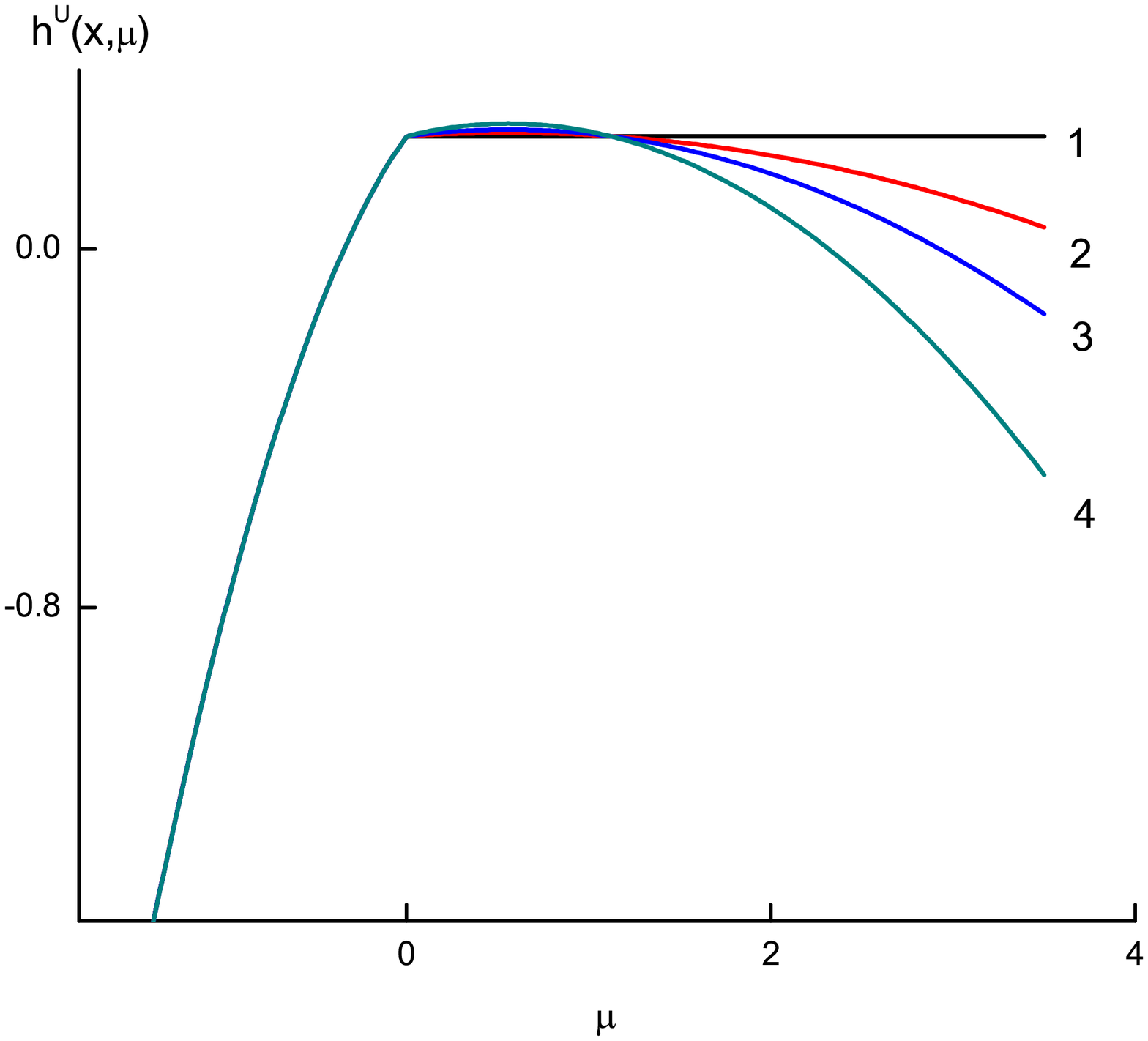}
\end{center}
\begin{center}
{Fig. 3. Distribution function in problem about weak evaporation.
Curves $1,2,3,4$ correspond to values $x=0,0.05,0.1,0.2$.}
\end{center}
\end{figure}
\clearpage
\medskip
{\sc Remark 4.2.}
Expansion (4.1) contains solutions of two
problems --- solution of problem about temperature jump
$$
\dfrac{h^T(x,\mu)}{g_T}=e^{-\gamma_0x}h^*(\mu)+h_{as}^T(x,\mu)
$$
and solution of the problem about weak evaporation
$$
\dfrac{h^U(x,\mu)}{g_T}=-e^{-\gamma_0x}h^*(\mu)+h_{as}^U(x,\mu).
$$

In these equalities are entered designations
$$
h^*(\mu)=\Big(\mu-\dfrac{2\mu^2}{\sqrt{5}}\Big)\dfrac{1+\gamma_0\si}
{\sqrt{\pi}(1+\gamma_0)},
$$
$$
h_{as}^T(x,\mu)=-\Big(1-\dfrac{1}{4\gamma_0}\Big)-\dfrac{\mu}{\sqrt{\pi}}+
\Big(1+\dfrac{1}{2\gamma_0}\Big)\Big(\mu^2-\dfrac{1}{2}\Big)+$$$$+
(x-\si)\Big(\mu^2-\dfrac{3}{2}\Big),
$$
$$
h_{as}^U(x,\mu)=\dfrac{1}{4\gamma_0}+\dfrac{\mu}{\sqrt{\pi}}-
\dfrac{\mu^2}{2\gamma_0}.
$$

{\sc Remark 4.3.} From resulted above equalities it is visible,
that dis\-tri\-bution function in the problem about temperature
jump is discontinuous in a point $ \mu=0$, and distribution
function in the problem about weak evaporation is continuous on
all real axis, including in the point $ \mu=0$.

Really, in case of the problem about temperature jump for
right-hand and left-hand limits in the point $ \mu=0$ we have
$$
h^{T}_+(x,+0)=g_T(x-1)\Big(-\dfrac{3}{2}\Big),
$$
$$
h^{T}_-(x,-0)=g_T(x+1)\Big(-\dfrac{3}{2}\Big).
$$

Hence, quantity of jump of distribution function of the reflected
from a wall and molecules flying to a wall in
the point $ \mu=0$ it is equal
$$
h^{T}_+(x,+0)-h^{T}_-(x,-0)=3g_T.
$$

Let us notice, that the quantity of this jump does not depend from
spatial variable $x$, i.e. it is identical at all
$x>0$.

In case of the problem about weak evaporation it is obvious, that
$$
h^{U}_+(x,+0)-h^{U}_-(x,-0)=0.
$$

{\sc Remark 4.4.} Let us notice, that in the problem about weak
eva\-po\-ration distribution function can be transformed to the
following form
$$
\dfrac{h^U(x,\mu)}{2U}=\dfrac{1}{4\gamma_0}+\dfrac{1}{\sqrt{\pi}}\Big(\mu-
\dfrac{2\mu^2}{\sqrt{5}}\Big)\Big[1-\dfrac{1+\gamma_0\si}
{1+\gamma_0}e^{-\gamma_0x}\Big].
$$
\medskip

From this formula, in particular, follows, that on border $x=0$
distribution  function of the reflected molecules is constant
(see fig. 2)
$$
\dfrac{h^U_+(0,\mu)}{2U}=\dfrac{1}{4\gamma_0}, \qquad \mu>0.
$$
\medskip

\begin{center}
  {\bf 5. Temperature jump and weak evaporation (condensation).
Distribution of macroparameters of gas}
\end{center}

Numerical calculations of coefficients of temperature and concentration
jump result in the following
$$
\varepsilon_T=1.5046g_T-0.5046(2U),
$$
$$
\varepsilon_n=-0.7477g_T-0.2523(2U).
$$

For comparison we will bring coefficients of temperature and
concen\-tra\-tion jumps found by means of the one-dimensional
kinetic equations with constant frequency of collisions
\cite{44}
$$
\varepsilon_T=1.3068g_T-0.4443(2U),
$$
$$
\varepsilon_n=-3.3207g_T-0.8958(2U).
$$

Let us consider distribution of concentration, mass velocity and
tem\-pe\-ra\-ture depending on coordinate $x $.

Let us begin with concentration distribution (numerical density). On
to definition it is received
$$
\dfrac{\delta n(x)}{n_0}=
\dfrac{1}{\sqrt{\pi}}\int\limits_{-\infty}^{\infty}
e^{-\mu^2}h(x,\mu)d\mu=$$$$=\varepsilon_n-g_Tx-
\dfrac{\dfrac{\gamma_0}{\sqrt{\pi}}-
\dfrac{\sqrt{\pi}}{4\gamma_0}}{\sqrt{\pi}(1+\gamma_0)}e^{-\gamma_0x}=
$$
$$
=\varepsilon_n-g_Tx-0.0317(2U-g_T)e^{-\gamma_0x}=$$$$
=-g_TN_T(x)-(2U)N_U(x),
$$
where
$$
N_T(x)=0.7477+x-0.0317e^{-\gamma_0x},
$$
$$
N_U(x)=0.2523+0.0317e^{-\gamma_0x}.
$$

Distribution of mass velocity at $x> 0$ is trivial
$$
U(x) \equiv U.
$$

Really, having taken advantage of the solution (4.1), easy
check up, that
$$
U(x)=\dfrac{1}{\sqrt{\pi}}\int\limits_{-\infty}^{\infty}e^{-\mu^2}
\mu h(x,\mu)d\mu \equiv U.
$$

This fact is quite obvious, and follows from the equation
continuity.

Let us consider temperature distribution. By definition we receive
$$
\dfrac{\delta T(x)}{T_0}=
\dfrac{2}{\sqrt{\pi}}\int\limits_{-\infty}^{\infty}e^{-\mu^2}\Big(
\mu^2-\dfrac{1}{2}\Big)h(x,\mu)d\mu=
$$
$$
=\varepsilon_T+g_Tx+
\dfrac{\dfrac{\sqrt{\pi}}{8}-\dfrac{1}{\sqrt{5}}}
{\sqrt{\pi}(1+\gamma_0)}e^{-\gamma_0x}(2U-g_T)=$$$$
=\varepsilon_T+g_Tx-0.0639e^{-\gamma_0x}(2U-g_T)=
$$$$=T_T(x)g_T-T_U(x)(2U),
$$
where
$$
T_T(x)=1.5046+x+0.0639e^{-\gamma_0x},
$$
$$
T_U(x)=0.5046+0.0639e^{-\gamma_0x}.
$$

\begin{figure}[t]
\begin{center}
\includegraphics[width=14cm,height=12cm]{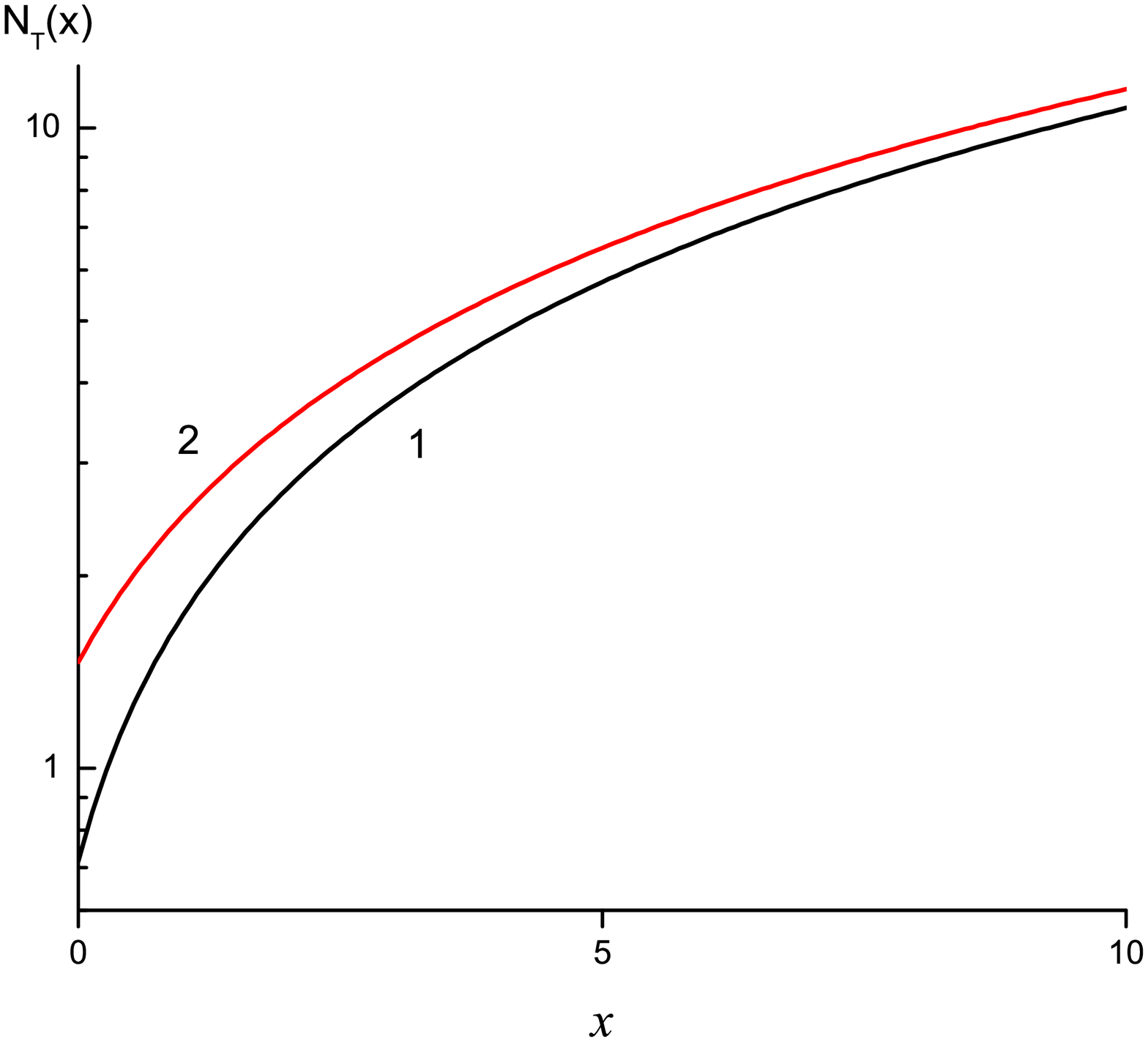}
\end{center}
\begin{center}
{Fig. 4. Behaviour of kinetic coefficient $N_T (x) $ (curve $1$),
The curve $2$ is the asymptotic $N_{Tas}(x)=0.7477+x$.}
\end{center}
\end{figure}

\begin{figure}[t]
\begin{center}
\includegraphics[width=14cm,height=12cm]{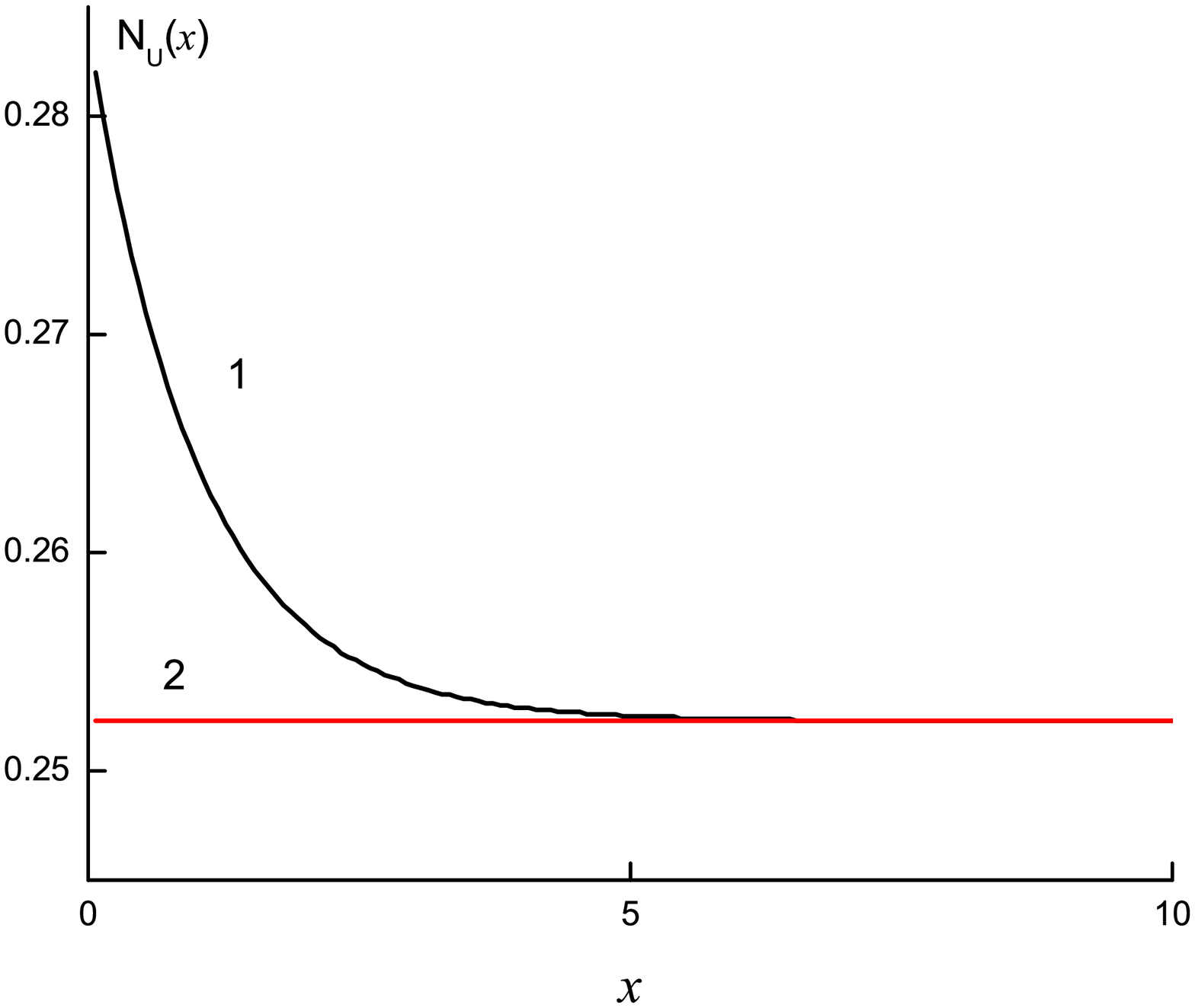}
\end{center}
\begin{center}
{Fig. 5. Behaviour of kinetic coefficient $N_T (x) $ (curve $1$),
The curve $2$ is the asymptotic  $N_{Uas}(x)=0.2523$.}
\end{center}
\end{figure}

\clearpage

\begin{figure}[t]
\begin{center}
\includegraphics[width=15cm,height=15cm]{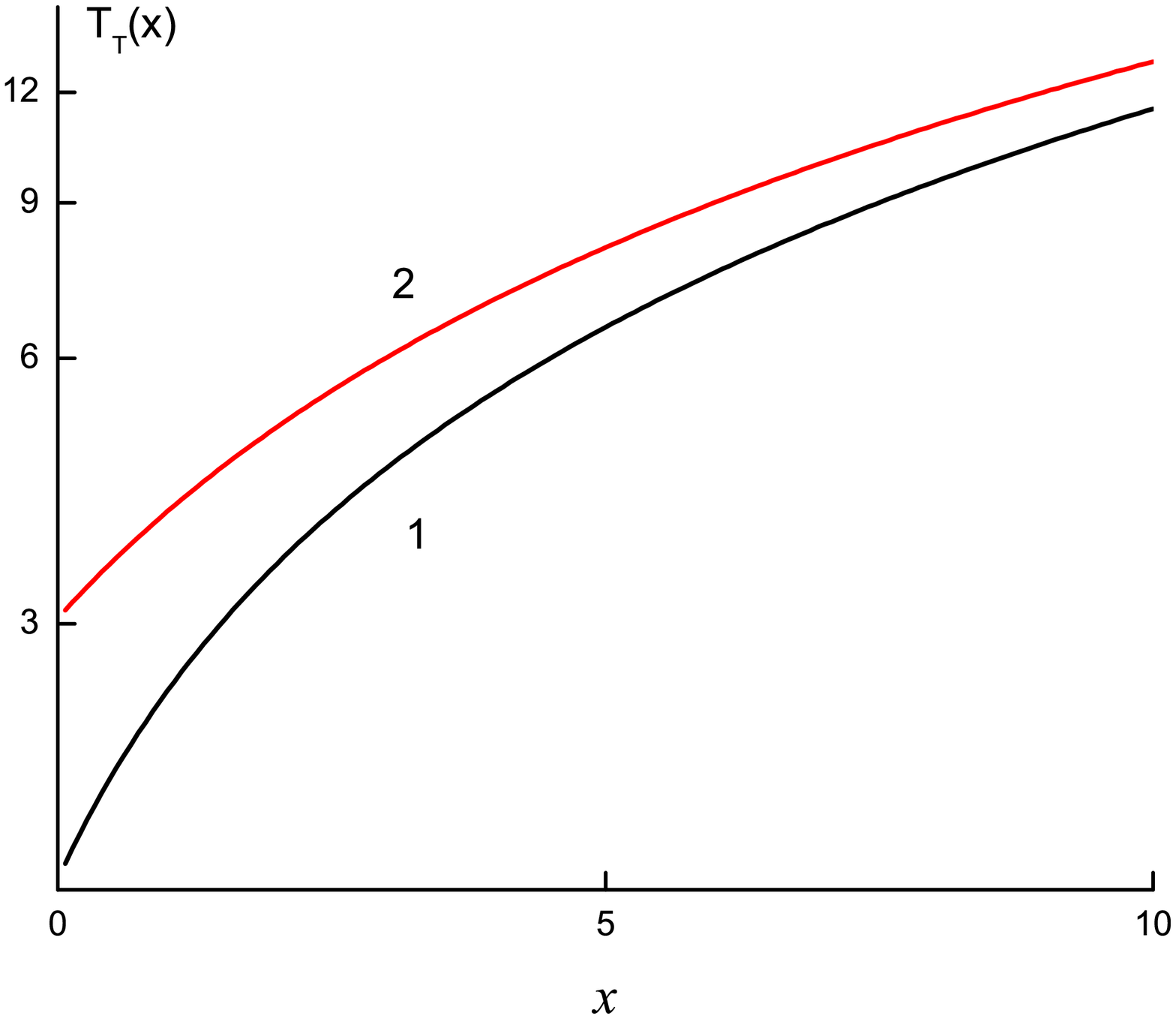}
\end{center}
\begin{center}
{Fig. 6. Behaviour of kinetic coefficient $N_T (x) $ (curve $1$),
The curve $2$ is the asymptotic  $T_{Tas}(x)=1.5046+x$.}
\end{center}
\end{figure}

\begin{figure}[t]
\begin{center}
\includegraphics[width=14cm,height=14cm]{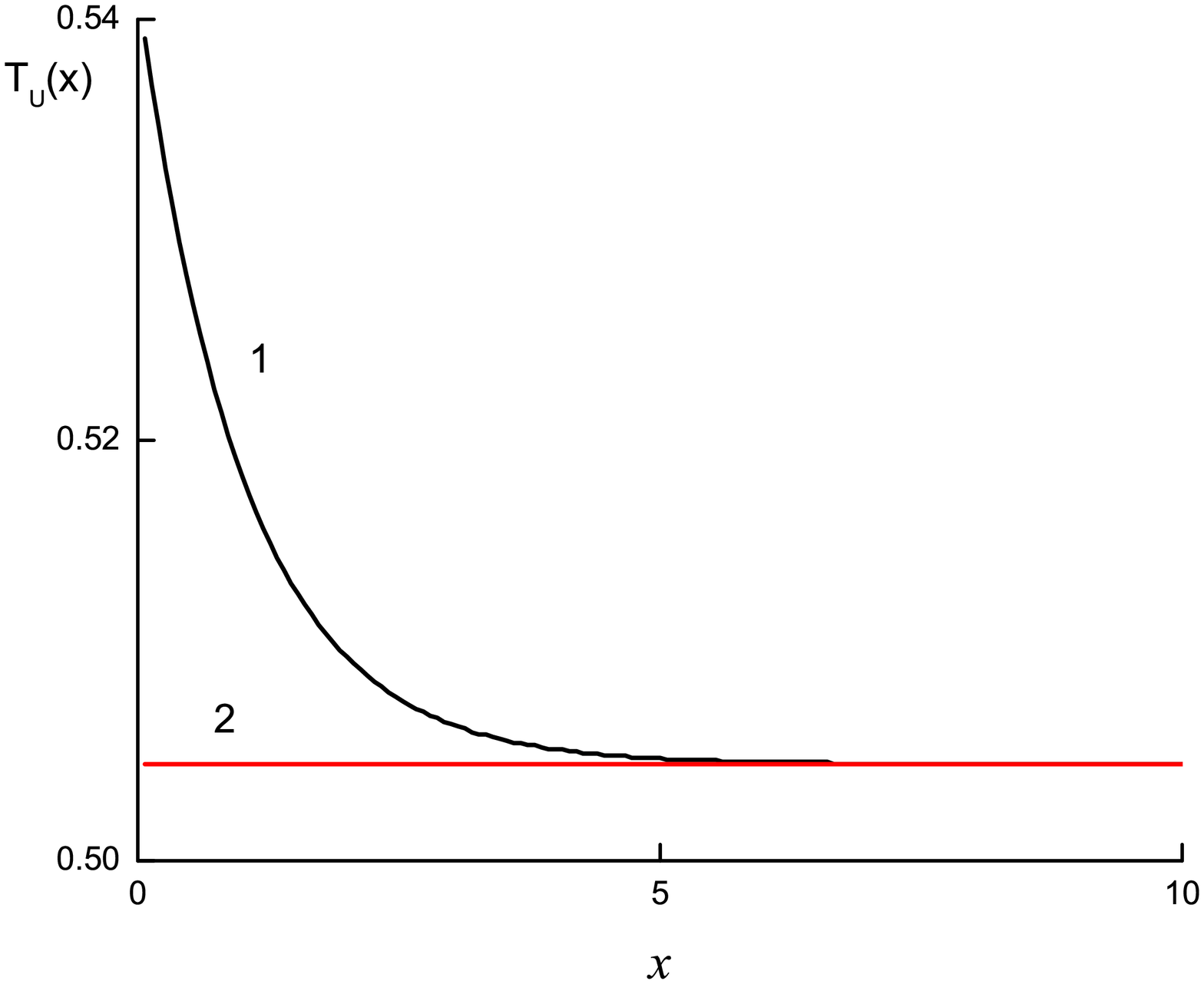}
\end{center}
\begin{center}
{Fig. 7. Behaviour of kinetic coefficient $N_T (x) $ (curve $1$),
The curve $2$ is the asymptotic  $T_{Uas}(x)=0.5046$.}
\end{center}
\end{figure}

\clearpage
\begin{center}
  \bf 6. Conclusion
\end{center}

In the present work the analytical solution of
boundary problems for the one-dimensional kinetic equation with
collisional frequency  of molecules proportional to the module
molecular velocity is considered.
This equation is the limiting case of affine
dependence of colliisional frequency of molecules on
the module of their velocity.

The analytical solution of generalized  Smolukhovsky problem
(about temperature jump and weak evaporation (condensation)) is considered.

Formulas for calculation of temperature and concentration
jumps are deduced.
Distribution function of gas molecules in explicit form, and
also distributions of concentration and temperature in
half-space $x>0$ are received.

It has appeared, that distribution function in
problem about tempe\-ra\-ture jump is discontinuous in the point
$ \mu=0$, and distribution function in problem about weak evaporation
is continuous at all velocities of molecules.
All necessary numerical calculations are done. It is spent
graphic research of distribution function of reflected molecules and
mole\-cules flying to the wall, and also all kinetic
coefficients are investigated.


\end{document}